# Uncovering the Origin of Divergence in the CsM(CrO$_4$)$_2$ (M = La, Pr, Nd, Sm, Eu; Am) Family through Examination of the Chemical Bonding in a Molecular Cluster and by Band Structure Analysis

Shane S. Galley,[1] Alexandra A. Arico,[1] Tsung-Han Lee,[2] Xiaoyu Deng,[2] Yong-Xin Yao,[3] Joseph M. Sperling,[1] Vanessa Proust,[1] Julia S. Storbeck,[1] Vladimir Dobrosavljevic,[4,5] Jennifer N. Neu,[5,6] Theo Siegrist,[5,6] Ryan E. Baumbach,[5] Thomas E. Albrecht-Schmitt[1,*] Nikolas Kaltsoyannis,[7,*] Nicola Lanatà[5,*]

[1]Department of Chemistry and Biochemistry, Florida State University, Tallahassee, Florida 32306, United States

[2]Department of Physics and Astronomy, Rutgers University, Piscataway, New Jersey 08856-8019, United States

[3]Department of Physics and Astronomy, and Ames Laboratory, U.S. Department of Energy, Iowa State University, Ames, Iowa 50011, United States

[4]Department of Physics, Florida State University, Tallahassee, Florida 32306, United States

[5]National High Magnetic Field Laboratory, Tallahassee, Florida, 32310, United States

[6]Department of Chemical Engineering, Florida State University, Tallahassee, FL 32310, United States

[7]School of Chemistry, The University of Manchester, Oxford Road, Manchester, M13 9PL, UK

*albrecht-schmitt@chem.fsu.edu

*nikolas.kaltsoyannis@manchester.ac.uk

*lanata@magnet.fsu.edu

**Abstract**

A series of *f*-block chromates, $CsM(CrO_4)_2$ (M = La, Pr, Nd, Sm, Eu; Am), were prepared revealing notable differences between the $Am^{III}$ derivatives and their lanthanide analogs. While all compounds form similar layered structures, the americium compound exhibits polymorphism and adopts both a structure isomorphous with the early lanthanides as well as one that possesses lower symmetry. Both polymorphs are dark red and possess band gaps that are smaller than the $Ln^{III}$ compounds. In order to probe the origin of these differences, the electronic structure of $\alpha$-$CsSm(CrO_4)_2$, $\alpha$-$CsEu(CrO_4)_2$, and $\alpha$-$CsAm(CrO_4)_2$ were studied using both a molecular cluster approach featuring hybrid density functional theory and QTAIM analysis, and by the periodic LDA+GA and LDA+DMFT methods. Notably, the covalent contributions to bonding by the *f* orbitals was found to be more than twice as large in the $Am^{III}$ chromate than in the $Sm^{III}$ and $Eu^{III}$ compounds, and even larger in magnitude than the Am-5*f* spin-orbit splitting in this system. Our analysis indicates also that the Am–O covalency in $\alpha$-$CsAm(CrO_4)_2$ is driven by the degeneracy of the 5*f* and 2*p* orbitals, and not by orbital overlap.

**Introduction**

Actinides beyond plutonium often have 5*f* electrons that are largely localized as evidenced by the superconducting behavior of americium metal.[1,2] In contrast, earlier actinides from at least uranium to plutonium display itinerant 5*f* electron behavior in their metallic states that extends to molecules where hybridization of 5*f* orbitals with ligand orbitals and delocalization of 5*f* electrons can occur.[3-9] This situation is further complicated by several factors that include the near degeneracy and greater radial extension of empty 6*d* orbitals, additional frontier orbitals coming into play (6*p*, 7*s*, and 7*p*), and reorganization of all of these orbitals upon complexation.[8-13] Relativistic effects and spin-orbit coupling (SOC) dominate the electronic structure in these heavy elements,[14-19] and the magnitude of crystal- and ligand-field splitting lies between that found in the 4*f* series and 5*d* transition metals. This situation is often termed the intermediate coupling regime.[14] Taken together, understanding the chemistry and physics of 5*f* elements represents the outer limits of current experimental and theoretical approaches. These challenges must be undertaken nevertheless for fundamental reasons that include understanding the evolution of electronic structure across the periodic table, and for practical applications, such as mitigating the environmental effects of the Cold War and improving the utilization of nuclear energy.

There are radiologic and reaction-scale challenges that are inherent to working with actinides that lie beyond uranium that often force the use of benign analogs for these elements. Examples of this include replacing $Pu^{IV}$ with $Ce^{IV}$ or $Am^{III}$ with $Eu^{III}$.[20-29] These substitutions are often based on similarities between ionic radii.[20] However, the aforementioned changes in electronic structure and the increased involvement of frontier orbitals in the actinides creates dissimilarities between the 4*f* and 5*f* series that manifests in unexpected coordination chemistry, electronic properties, and reactivity. For example, the reactivity and coordination environments of cerium and plutonium diverge in phosphonates,[23,25] carboxypyridinonates,[30,31] and hydroxypyridinonates.[21,32-34] The enthalpy of complexation of Am[III] by softer donors ligands is notably stronger than it is for Eu[III] and can be exploited for separating Am[III] from lanthanides in used nuclear fuel recycling.[10,35-40] Contracted M-L bonds have also been measured and calculated in M[N(EPR$_2$)$_2$]$_3$ complexes (M = U, Pu; E = S, Se, Te; R = Ph, *i*Pr, H) that are consistent with enhanced covalency in An–E bonds versus that found with lanthanides.[41]

Variance occurs not only between the 4*f* and 5*f* series, but also between early and late actinides.[19,42-50] Recent studies on the reduction of An[III] cyclopentadienyl complexes to An[II] have shown bifurcation in the ground states of the resultant species with U[II] existing in a 5$f^n$6$d^1$ (5$f^3$6$d^1$) state; whereas Pu[II] adopts a 5$f^{n+1}$6$d^0$ (5$f^6$) configuration.[51-53] These differences in bonding between actinides are further illustrated by U[IV] and Pu[IV] β-ketoiminates where contributions from both 7*s* and 6*d* orbitals were found in the U–O bonds; whereas the only metal-based orbitals participating in the Pu–O bonds are the 6*d* orbitals.[11] Rare examples of studies on the complexation of Bk[III] and Cf[III] have shown that the more negative bond enthalpies are the result of increased covalency, and that part of the origin of this effect is driven by the degeneracy of actinide 5*f* orbitals and ligand orbitals.[44,46,55-57]

While many of the aforementioned examples have been pursued in order to provide a basic understanding of structure and bonding in *f*-element compounds, some of these materials are of practical importance and play roles in mitigating the environmental legacy of the Cold War. Among the components of nuclear waste of particular concern are large amounts of the mutagen, chromate, that is present in waste tanks because of its use in antiquated separations methods such as REDOX,[58] and as a corrosion inhibitor for the tanks themselves. Complicating matters further, chromates also form undesirable inclusions in the form of spinels during vitrification of nuclear waste.[59] Th[IV] and U[VI] chromates have been the subject of numerous investigations and show a vast array of structural topologies.[60-66] However, both of these actinide cations are $5f^0$, and therefore lack many of the interesting electronic characteristics found in later actinide compounds.

In contrast to thorium and uranium compounds, transuranium chromates are largely unexplored, and most examples that appear in the literature are poorly characterized. Among the few well-characterized compounds is actually a rare example of a $Am^V$ compound, $Cs_3AmO_2(Cr_2O_7)_2 \cdot H_2O$, that was obtained via ozonation of $Am^{III}$.[67,68] The most stable oxidation state of americium is III+ and is likely more relevant to americium speciation in tank waste. However, an $Am^{III}$ chromate has yet to be reported. In order to address this issue, we have undertaken the investigation of the synthesis, structure, and properties of $Am^{III}$ chromates. These results are placed within the context of other trivalent *f*-element chromates by completing a comprehensive study of the $Ln^{III}$ compounds that form under the same conditions.[69-71] Here we examine the $CsM(CrO_4)_2$ (Ln = La, Pr, Nd, Sm, Eu; Am) family of compounds and show that $Am^{III}$ can form the same structure type as found with $Ln^{III}$ ions that possess similar ionic radii as well as a structure type not yet observed with lanthanides. We also show that the band gap for $CsAm(CrO_4)_2$ is smaller than that observed for the lanthanides; necessitating an examination of the electronic structure of these compounds. The bonding is probed computationally by both excising a cluster that describes the local coordination environment and via band-structure calculations.

**Experimental Section**

**Syntheses**. *Caution!* $^{243}Am$ *($t_{1/2}$ = 7370 y) is an intense α emitter and also emits penetrating γ-rays up to 142 KeV in energy. Of equal importance, its α-decay product is $^{239}Np$, which is both short lived ($t_{1/2}$ = 2.35 d) and an even more potent γ-ray emitter with energies up to 278 KeV. Studies on the bulk synthesis of americium compounds can only be conducted in an appropriately equipped* *radiologic facility. In this case, all studies were performed in a Category II nuclear hazard facility. The manipulations of solids were carried out using negative-pressure gloveboxes with a series of filters being used to remove any particulate matter that could hypothetically exit the glovebox exhaust. The air in the lab is also heavily filtered through a series of eight filters that include carbon and HEPA filters. When samples are removed from the gloveboxes the researchers wear respirators and air sampling is performed to ensure that radioactive particulates have not been aerosolized. Owing to the γ emission from $^{243}$Am and its daughters, researchers are also shielded using moveable lead walls that are placed in front of gloveboxes, lead wells, lead brick walls, thick lead sheets around furnaces, and long lead vests are worn to protect researchers.* $^{243}$AmO$_2$ (98% purity) was obtained from Oak Ridge National Laboratory. Ln(NO$_3$)$_3$·6H$_2$O (La, Pr, Nd, Sm, Gd) (99.9%; Sigma-Aldrich), Ln$_2$O$_3$ (Eu, Dy, Ho, Er, Tm, Yb, Lu) (99.9%; Sigma-Aldrich), Tb$_4$O$_7$ (99.9%; Sigma-Aldrich), Cs$_2$CrO$_4$ (99.5%; Sigma-Aldrich), and Cs$_2$Cr$_2$O$_7$ (99.5% Sigma-Aldrich) were used as received.

CsLn(CrO$_4$)$_2$ (Ln = La, Pr, Nd, Sm) were prepared by loading 0.2 mmol of Ln(NO$_3$)$_3$·6H$_2$O (Ln = La, Pr, Nd, or Sm), 0.075 mmol of Cs$_2$CrO$_4$, and 0.075 mmol of Cs$_2$Cr$_2$O$_7$, and 2 mL of H$_2$O into a 23 mL PTFE-lined autoclave. The autoclaves were sealed and heated in a box furnace at 200 °C for 48 hours with a 48-hour cooling period following thereafter. Crystals of all compounds were isolated directly from the hydrothermal reactions and rinsed with water to remove excess chromates. Large, well-faceted gold or yellow-orange offset prisms were isolated for CsLa(CrO$_4$)$_2$ and CsPr(CrO$_4$)$_2$, respectively. The crystals of CsNd(CrO$_4$)$_2$ and CsSm(CrO$_4$)$_2$ form green columns and gold plates, respectively. These reactions were then scaled down to appropriate levels for work with $^{243}$Am. The reactions were carried out again to ensure that crystal growth still occurs. Once this was verified, the work with americium was conducted. Reactions were performed on the exact scale described below for all of the lanthanides described in this work. It should also be noted that reactions were also carried out with Eu through Lu. These reactions result in the formation of Ln(OH)(CrO$_4$) (Ln = Eu, Gd, Tb, Dy, Ho, Er, Tm, Yb, or Lu). Europium represents the crossover in this system and the europium reactions appears to contain a mixture of products. Studies on these compounds will be reported elsewhere. Reactions with Ce$^{III}$ starting materials result in an immediate redox reaction prior to heating with a corresponding color change to black and formation of amorphous black and green solids. Alternatively, all of these compounds can be prepared by reactions between Cs$_2$CrO$_4$ and Ln(NO$_3$)$_3$·nH$_2$O in 2:1 ratio using the same hydrothermal conditions described above. This results in improvement in crystal quality, purity, and yield. Typical yields are ~70% based on the lanthanide content. Furthermore, this latter method allows one to isolate pure β-CsLn(CrO$_4$)$_2$ (Ln = Nd, Sm, Eu, Gd, Tb, Dy, and Ho); whereas using the former conditions the products are mixtures that also contain the aforementioned Ln(OH)(CrO$_4$) compound beginning at europium. All in cases except for Am, SEM-EDS data were obtained that help to confirm the 1:1:2 ratio of Cs:Ln:Cr.

**α-CsAm(CrO$_4$)$_2$ and β-CsAm(CrO$_4$)$_2$.** Two different polymorphs of CsAm(CrO$_4$)$_2$ can be prepared using different synthetic conditions. α-CsAm(CrO$_4$)$_2$ is prepared by first synthesizing Am(NO$_3$)$_3$·$n$H$_2$O from AmO$_2$. Am(NO$_3$)$_3$·$n$H$_2$O forms by reacting multiple 100 μL aliquots of 5 M HNO$_3$ with 0.02 mmol of AmO$_2$ and slowly fuming the mixture to dryness. A color change from black to light yellow is indicative of reduction to Am$^{III}$, and this putative Am$^{III}$ nitrate along with 0.01 mmol of Cs$_2$CrO$_4$ and 0.01 mmol of Cs$_2$Cr$_2$O$_7$, and 200 μL of water were loaded into the 10 mL PTFE-lined autoclave. These reactions were also carried out using the corresponding Ln starting materials. For Nd and Sm, these reactions results in the formation of β-CsLn(CrO$_4$)$_2$; the α form as not yet been obtained with any lanthanide. The autoclave was sealed and heated in a box furnace at 200 °C for 48 hours with a 48-hour cooling period. Very dark red (nearly black) rods ~200 μm in length were isolated directly from the mother liquor. β-CsAm(CrO$_4$)$_2$ was instead prepared with a hydrous AmCl$_3$ starting material on a larger scale. A 24-hour digestion of 0.0358 mmol of $AmO_2$ with 500 μL of 5 M HCl at 150 °C yielded a solution of $Am^{III}$. This solution was fumed to dryness with a heat lamp, after which 0.0179 mmol of $Cs_2CrO_4$ and 0.0179 mmol of $Cs_2Cr_2O_7$ and 500 μL of water were loaded into a 10 mL autoclave. The same heating profile was followed as used in the synthesis of α-$CsAm(CrO_4)_2$. Large, dark-red blocks of β-$CsAm(CrO_4)_2$ were isolated and rinsed with water.

**Crystallographic Studies.** Single crystals of all compounds were adhered to Mitogen loops with immersion oil and then mounted on a goniometer under a cold stream set at 100 K. The crystals were then optically aligned on a Bruker D8 Quest X-ray diffractometer using a digital camera. Diffraction data were obtained by irradiating the crystals with an IμS X-ray source (Mo Kα, $\lambda$ = 0.71073 Å) with high-brilliance and high-performance focusing multilayered optics. Bruker software was used for determination of the unit cells, data collection, and integration of the data. Lorentz, polarization, and absorption corrections were also applied. A hemisphere of data was collected for all crystals. The structures were solved by direct methods and refined on $F^2$ by full-matrix least-squares techniques using the program suite SHELXTL.[72] Structure factors for americium were not present in the SHELX software at the time of these studies were performed, but were recently added by G. M. Sheldrick. Thus a new SFAC command had to be added to the instructions file that defines the scattering factors for americium. This procedure no longer has to be done if one uses the most recent version of SHELX. Some of these compounds crystallize in less common space groups and the solutions were checked for missed symmetry using PLATON.[73] The Crystallographic Information Files (CIF) are available from the Cambridge Crystal Structure Database Center: 1571937 (α-$CsAm(CrO_4)_2$), 1571730 (β-$CsAm(CrO_4)_2$), 1571731 ($CsLa(CrO_4)_2$), 1571733 ($CsNd(CrO_4)_2$), 1571732 ($CsPr(CrO_4)_2$), 1571734 ($CsSm(CrO_4)_2$), and 1589766 (β-$CsEu(CrO_4)_2$). Selected

crystallographic data are provided in Table S1. We have simplified the formula to CsM(CrO$_4$)$_2$. In some cases, the formulae are more correctly (crystallographically) expressed as Cs$_2$M$_2$(CrO$_4$)$_4$.

**UV−Vis-NIR Spectroscopy.** Single crystals of each compound were placed on quartz slides under medium-viscosity Krytox oil. A Craic Technologies microspectrophotometer was used to collect optical data in the UV-vis-NIR region. Irradiation of the samples was performed with a mercury light source. Absorption spectra of β-CsSm(CrO$_4$)$_2$ and β-CsEu(CrO$_4$)$_2$ are provide in Figure S1. These data reveal band gaps from these materials of 2.462 and 2.435 eV for β-CsSm(CrO$_4$)$_2$ and β-CsEu(CrO$_4$)$_2$, respectively.

**Magnetic Susceptibility Measurements.** Magnetism measurements were performed on polycrystalline samples of α-CsAm(CrO$_4$)$_2$, β-CsNd(CrO$_4$)$_2$, and β-CsEu(CrO$_4$)$_2$ using a Quantum Design MPMS under an applied field of 10 kOe for 4 K < 300 K. Plots of these data for β-CsNd(CrO$_4$)$_2$, and β-CsEu(CrO$_4$)$_2$ are provided in the Figure S2. The americium sample was sealed inside two, different, custom-built Teflon capsules. The first capsule has a piston design and fits inside of the second capsule that screws closed and is also taped to ensure that it cannot open during data collection. Datasets were collected with the capsules both empty and full, and the background from the sample holder was subtracted from the signal. Appropriate diamagnetic corrections were also applied. Data for α-CsAm(CrO$_4$)$_2$ are not provided because the sample was nonmagnetic over the entire temperature range studied.

For β-CsSm(CrO$_4$)$_2$, and β-CsEu(CrO$_4$)$_2$ temperature dependent magnetic susceptibility $\chi = M/H$ were obtained from randomly-oriented powders loaded in gelatin capsules, where $M$ is the magnetization under an applied magnetic field $H$ = 1000 Oe. (a, b, c) Figure S1 shows the temperature dependences of $\chi$, $\chi T$, and $\chi^{-1}$, respectively. As shown in panel a, the curves for both compounds are weakly temperature dependent on the range 25 – 300 K. In panel b we show $\chi T$, which will approach the Curie-constant in the high temperature limit if Curie-Weiss behavior occurs ($\chi T \approx 0.09$ cm$^3$K/mol for Sm$^{III}$ and 0 for Eu$^{III}$). At 300 K we find (i) that for the Sm$^{III}$ version $\chi T$ is both non-saturating and larger than expected and (ii) for the Eu$^{III}$ version it is positive and non-saturating. Furthermore, in panel c we show that for both compounds there is no region that obviously exhibits Curie-Weiss behavior. This indicates that for both compounds there is a sizable paramagnetic signal $\chi_0$ that is only weakly temperature dependent. The case of Sm$^{III}$ might be more complex because these ions are expected to carry a small localized magnetic moment. However, it is likely that the crystal electric field splitting for the Sm$^{III}$ Hund's rule multiplet interrupts a simple Curie-Weiss behavior on the temperature scale that is presented. As a result, the ligand field again dominates $\chi$.

**Cluster calculations.** DFT calculations were performed on a cluster representation of α-CsAm(CrO$_4$)$_2$ and α-CsEu(CrO$_4$)$_2$ (which cannot be prepared by the synthetic methods reported herein) as discussed in the main text. The Gaussian 09 code, revision D.01 was employed.[74] (14s 13p 10d 8f 5g)/[10s 9p 5d 4f 3g] segmented valence basis sets were used for Am and Eu, with Stuttgart-Bonn variety, 60- and 28-electron relativistic pseudopotentials respectively,[75,76] and a (7s 6p)/[5s 4p] basis set plus the 46-electron relativistic pseudopotential for Cs.[75] Dunning's cc-pVTZ basis set was used for Cr and O. The PBE0 functional[77] was used in conjunction with the ultrafine integration grid. The SCF convergence criterion was set to 10$^{-6}$, and the geometry convergence criterion was relaxed slightly from the default using iop(1/7=667) that produces 10$^{-3}$ au for the maximum force.

QTAIM analyses were performed using the AIMALL program package,[78] with .wfx files generated in Gaussian 09 used as input. NBO analyses were performed with the NBO6 code.[79]

**Band-Structure Calculations of α-CsAm(CrO$_4$)$_2$.** The electronic structure of α-CsAm(CrO$_4$)$_2$ was investigated utilizing the Local Density Approximation (LDA)[80] in combination with dynamical mean field theory (LDA+DMFT)[81-83] and in combination with the Gutzwiller

Approximation (LDA+GA).[84-86] Both of these computational approaches are powerful tools widely used to study strongly-correlated electron systems that enables us to take into account the strong Am-5$f$ electron correlations in α-CsAm(CrO$_4$)$_2$. We utilize the DFT code WIEN2K,[87-89] and employ the standard fully-localized limit form for the double-counting functional. The LAPW interface between LDA and DMFT/GA employed in our calculations was implemented as described in Ref. [89]. The LDA+DMFT simulations are performed utilizing the Continuous Time Quantum Monte Carlo (CTQMC) impurity solver[90] at $T = 290\ K$. Consistently with previous work,[86] we assumed that the Hund's coupling constant is $J = 0.7\ eV$ and that the value of the screened Coulomb interaction strength is $U = 6.0\ eV$. Since our experiments have all been performed above the Neél temperature of the system, in our simulations we assume from the onset a paramagnetic wavefunction, i.e., a solution that does not spontaneously break the symmetry of the system. Spin-orbit coupling is fully taken into account in our calculations.

**Results and Discussion**

**Structure and Topological Analysis.** CsLn(CrO$_4$)$_2$ (Ln = La, Pr) crystallize in the monoclinic space group $P2_1/c$ and form layered structures. The layers are composed of lanthanide chromate chains that propagate along the $2_1$ screw parallel to the $b$ axis. These chains comprise CrO$_4^{2-}$ tetrahedra that corner- and edge-share with Ln$^{III}$ polyhedra. Ln$^{III}$ cations bridge between the chains creating layers that extend parallel to the [$bc$] plane. Cs$^+$ cations fill the interlayer space as illustrated in Figure 1. RbLa(CrO$_4$)$_2$, RbPr(CrO$_4$)$_2$, and KLa(CrO$_4$)$_2$ are isotypic with these compounds and have been previously reported.[69]

The Ln$^{III}$ centers are nine-coordinate with an approximate muffin geometry in CsLn(CrO$_4$)$_2$ (Ln = La, Pr).[91] The coordination environment contains nine oxygen atoms donated from CrO$_4^{2-}$ units, as shown in Figure 2a. For CsLa(CrO$_4$)$_2$, the La–O bond distances range from 2.474(4) to 2.589(4) Å. The lanthanide contraction is observed in this system with the Pr–O distances ranging from 2.422(3) to 2.558(3) Å. There are two crystallographically unique Cr$^{VI}$ sites with Cr–O distances that occur from 1.605(3) to 1.682(3) Å. The CrO$_4^{2-}$ tetrahedra are distorted, as evidenced by both these variable bond distances and non-ideal bond angles that range from 100.48(15)° to 112.84(17)°. Ln–O and Cr–O bond distances are provided in Table S1.

Gradual symmetry reduction and ultimately collapse of the layers occurs as the Ln$^{III}$ ionic radii diminish. Beginning at Nd$^{III}$ the crystallographic symmetry is lowered to *P*2/*c*. However, the structures remain quite similar as shown in Figure 3. The rubidium analogs have been previously reported, but do not adopt the same structure type as those reported here.[92] As the ionic radii continue to contract the Ln$^{III}$ coordination number decreases from nine to eight, and Nd$^{III}$, Sm$^{III}$, and Eu$^{III}$ cations are found within LnO$_8$ trigonal dodecahedra.[21] A view of the local coordination environment around these cations is shown in Figure 2b. For β-CsNd(CrO$_4$)$_2$, the Nd–O bond distances range from 2.381(2) to 2.573(2) Å. In accordance with the lanthanide contraction, slightly shorter Sm–O bond distances are observed in CsSm(CrO$_4$)$_2$ and occur from 2.343(3) to 2.546(3) Å. The reason why these contractions are so obvious even with different data collection temperatures is that lanthanides are skipped between La$^{III}$ and Pr$^{III}$ and between Nd$^{III}$ and Sm$^{III}$. In the former case, cerium is absent because in the presence of chromate it oxidizes to Ce$^{IV}$, and in the latter case promethium is radioactive with no long-lived isotopes. The Nd$^{III}$ and Sm$^{III}$ compounds contain only one crystallographically unique CrO$_4^{2-}$ site and greater variance in Cr–O bond distances is observed in these structures when compared to that observed for La$^{III}$ and Pr$^{III}$.[69,94-97] The Cr–O bond distances in β-CsNd(CrO$_4$)$_2$ and β-CsSm(CrO$_4$)$_2$ range from 1.616(2) to 1.692(2) Å and 1.609(3) to 1.692(2) Å, respectively. Again the $CrO_4^{2-}$ tetrahedra are distorted with bond angles ranging from 101.42(13) to 112.61(13)°. Bond distances for β-CsNd(CrO$_4$)$_2$, β-CsSm(CrO$_4$)$_2$, and β-CsSm(CrO$_4$)$_2$ are provided in Table S2.

Polymorphism in α-CsAm(CrO$_4$)$_2$ and β-CsAm(CrO$_4$)$_2$ is likely representative of small energetic differences between different structure types in this system. Similar polymorphism is also observed in M(IO$_3$)$_3$ (M = La – Lu, Am, Cm, Bk, Cf).[98-108] α-CsAm(CrO$_4$)$_2$ is triclinic and not isotypic with any of the other compounds described in this work. In contrast, β-CsAm(CrO$_4$)$_2$ is isomorphous with β-CsLn(CrO$_4$)$_2$ (Ln = Nd, Sm, Eu). The ionic radius of Am$^{III}$ most closely matches that of Nd$^{III}$, and this latter result is expected.[20] Even though α-CsAm(CrO$_4$)$_2$ possesses lower symmetry than β-CsAm(CrO$_4$)$_2$, both structures contain one crystallographically unique Am$^{III}$ site. However, the reduced symmetry of α-CsAm(CrO$_4$)$_2$ does give rise to two crystallographically-unique chromium centers rather than one, and this alters the topology of the layers from that observed in β-CsAm(CrO$_4$)$_2$.

The layers in β-CsAm(CrO$_4$)$_2$ have already been described because they are isomorphous with the β-CsLn(CrO$_4$)$_2$ (Ln = Nd, Sm, Eu) compounds. Thus, we will only detail those found in α-CsAm(CrO$_4$)$_2$. The layers in α-CsAm(CrO$_4$)$_2$ extend parallel to the [$ab$] plane and are composed of edge-sharing chains of Am$^{III}$ polyhedra connected by alternating corner- and edge-sharing $CrO_4^{2-}$ tetrahedra (Cr1) and strictly corner-sharing $CrO_4^{2-}$ tetrahedra (Cr2), as shown in Figure 4.

The Am$^{III}$ cations in both structure types are eight-coordinate with the americium site in α-CsAm(CrO$_4$)$_2$ being found within a polyhedron best approximated by a bicapped trigonal prism; whereas the site in β-CsAm(CrO$_4$)$_2$ is closer to a trigonal dodecahedron. While neither compound has high enough symmetry to have these idealized geometries around the metal centers, it should be noted that the former unit has formal $C_{2v}$ point symmetry, whereas the latter is $D_{2d}$. Thus, these compounds provide examples of higher crystallographic symmetry that possibly give rise to higher (approximate) point symmetry at the metal centers. Further elaboration of the local coordination reveals seven $CrO_4^{2-}$ anions binding the Am$^{III}$ centers in α-CsAm(CrO$_4$)$_2$ versus six in β-CsAm(CrO$_4$)$_2$ as shown in Figure 5. In other words, there are two chelating chromate anions in β-CsAm(CrO$_4$)$_2$ and only one in α-CsAm(CrO$_4$)$_2$.

Selected bond distances are provided in Table 1, and additional distances are given in the SI. Of particular importance are the statistically-equivalent, average Am–O bond lengths of 2.438(4) and 2.439(4) Å in α-CsAm(CrO$_4$)$_2$ and β-CsAm(CrO$_4$)$_2$, respectively. As previously mentioned, use of modern structural data has shown that the ionic radius of Am$^{III}$ most closely matches with that of Nd$^{III}$;[20] although it would be helpful to have high-resolution Pm$^{III}$ structures to place greater confidence in this assignment. A number of quests have been embarked on to find evidence of covalency via shortened An–L bonds versus those observed in Ln–L where the 5*f* and 4*f* ions chosen have similar ionic radii.[19,24,31,41,44,45] In some cases the differences between Ln–L and An–L bond distances are not statistically significant, whereas in the others a contraction that could be attributable to covalency is noted. M[N(EPR$_2$)$_2$]$_3$ complexes (M = U, Pu; E = S, Se, Te; R = Ph, *i*Pr, H) provide examples where shorter An–E bond lengths were observed than found with lanthanides of similar ionic radii.[41] Of equal significance, contraction of An$^{III}$–L bond distances from early actinides to post-curium elements is far from being monotonic as observed in An(Hdpa)$_3$ (An = Bk, Cf) complexes.[19,44,45] Based on these observations and the difficulties involved in clearly identifying statistically-significant differences between Ln–L and An-L bond distances, considerable time was invested in obtaining Nd$^{III}$ and Am$^{III}$ chromate structural data under as similar data collection conditions as possible, with temperature and high resolution being the most important factors. We first compared data obtained at 298 K to data collected at 130 K, and observed that the average bond distance change of the Nd–O bonds is on the order of 0.01 Å between these temperatures. A comparison of the average Nd–O bond distance with the average Am–O bond distance determined from diffraction data collected at 130 K reveals values of 2.450(2) and 2.439(4), respectively. While it appears that the average Am–O bond distance is shorter, this does not hold true at the 3σ limit.

**UV−Vis−NIR Spectroscopy**. Absorption data were collected for all compounds from single crystals using a microspectrophotometer. Characteristic *f-f* transitions for the trivalent lanthanides are observed where expected for some of the lanthanide compounds.[109] In some compounds, however, the *f-f* transitions are buried beneath, or simply obscured because of their low molar absorptivity, by the intense ligand-to-metal charge-transfer (LMCT) bands/semiconductor edge. This is the case for α-CsSm(CrO$_4$)$_2$ and β-CsEu(CrO$_4$)$_2$ as shown in Figure S2. Analysis of the Cs$_2$CrO$_4$ starting material reveals only a charge-transfer band extending to 450 nm. In contrast, the spectra of α-CsAm(CrO$_4$)$_2$ and β-CsAm(CrO$_4$)$_2$ display an intense absorption feature extending through the visible spectrum to around 720 nm that explains the dark red color of the crystals, as shown in Figure 6a. Characteristic intra-*f* transitions for Am$^{III}$ are also present, such as the $^7F_0 \rightarrow ^7F_6$ transition centered near 815 nm.[110] Such strong absorption across the visible range is likely indicative of semiconducting behavior. The absorbance is displayed also as a function of the optical energy via the Kubelka-Munk function in Figure 6b. The bandgaps for both α-CsAm(CrO$_4$)$_2$ and β-CsAm(CrO$_4$)$_2$ are approximately 1.65 eV, whereas the bandgaps for α-CsSm(CrO$_4$)$_2$ and β-CsEu(CrO$_4$)$_2$ are ~2.45 eV.

**Computational analysis of Am–O bonding in a Molecular Cluster.** In order to probe the nature of the Am–O bonding in α-CsAm(CrO$_4$)$_2$ we utilize first molecular quantum chemistry at the hybrid DFT level (PBE0). The system studied contains a single Am site surrounded by seven CrO$_4$ units; these atoms were fixed at their crystallographically determined positions. To balance the 11– charge this cluster carries, 11 Cs$^+$ counterions were added and their positions optimized. The neutral molecular cluster thus analyzed was Am(CrO$_4$)$_7$Cs$_{11}$, Cartesian atomic coordinates of which are given in the Supporting Information.

Over the last few years there has been much debate about the nature of covalency in the 5*f* series. Perturbation theory holds that, to first order, the mixing of molecular orbitals $\phi_i$ and $\phi_j$ is governed by the mixing coefficient $t_{ij}^{(1)}$:

$$t_{ij}^{(1)} \propto \frac{-H_{ij}}{e_i - e_j}, \qquad [1]$$

where the off diagonal elements of the Hamiltonian matrix $H_{ij}$ are related to the overlap between the orbitals, and the denominator is the difference between the corresponding energies. Thus, large orbital mixings can arise when $\phi_i$ and $\phi_j$ are close in energy, without there necessarily being significant spatial orbital overlap. The actinide community is now cognisant of the distinction between the more traditional *overlap-driven* covalency and *energy-driven* covalency that arises from the near degeneracy of metal and ligand orbitals.[109,110] The latter is common in the transuranic elements; as the actinide series is crossed the 5*f* orbitals become energetically stabilized and radially more contracted. Thus, at a certain point (dependent on the metal and the supporting ligand set) they become degenerate with the highest lying ligand based functions, yet are too contracted for there to be significant spatial overlap.

The atomic orbital (AO) contributions to the 20 highest occupied α spin canonical Kohn-Sham molecular orbitals (MOs) are shown in SI (Table S4). They are composed of Am *f* and oxygen *p* character, and there is extensive mixing of metal and ligand AOs in many MOs. This is illustrated in Figures 7 and 8 which show, respectively, MOs 266α and 262α. These images suggest that Am(CrO$_4$)$_7$Cs$_{11}$ is a good example of energy-driven covalency; there are many atomic orbital contributions to the two molecular orbitals shown, but little or no spatial overlap of the individual atom-centered orbitals. This conclusion is reinforced by Natural Bond Orbital (NBO) analysis. The NBO approach is an orbital localization procedure that attempts to recast the canonical Kohn-Sham orbital structure in terms of more chemically intuitive localized orbitals, emphasizing the Lewis-like molecular bonding pattern of electron pairs. Applying the technique to Am(CrO$_4$)$_7$Cs$_{11}$ yields no Am–O NBOs. Furthermore, the NBO calculation yields six α spin orbitals that are all greater than 99.9% Am 5*f* in character, i.e., they are the six unpaired 5*f* electrons expected for an Am$^{3+}$ center at the scalar relativistic level. This picture is very different from the delocalized nature of the Kohn-Sham orbitals, and suggests highly ionic Am–O bonding. In support of this, the Natural and Mulliken spin densities are 5.93 and 6.02 respectively (very close to the 6 expected for an Am$^{3+}$ ion). Furthermore, the expectation value of the $S^2$ operator is $\langle S^2 \rangle = 12.01$; a pure heptet state would have is $\langle S^2 \rangle = 12$. Hence, these results indicate essentially zero spin contamination in the wavefunction.

In principle, there is an infinite number of orbital representations we could choose to analyze. By contrast, the Quantum Theory of Atoms in Molecules (QTAIM) focuses not on orbitals but on the topology of the electron density, and allows us to analyze actinide covalency in an alternative way, ideally distinguishing energy-driven from overlap-driven effects; the former will not lead to a significant build-up of electron density in the internuclear region, while the latter should do so.[12,111-114] The QTAIM states that there is a bond critical point (BCP) between every two atoms bonded to each other, with the BCP located at the minimum in the electron density along the bond path, the line of maximum electron density between the two atoms. The values of the electron and energy densities $\rho$ and $H$, and $\nabla^2\rho$, at the BCP can be used in analyzing the nature of the bond. Large $\rho$ values are associated with covalent bonds, and $H$ is negative for interactions with sharing of electrons, with its magnitude indicating the covalency of the interaction. $\nabla^2\rho$ is also generally significantly less than zero for covalent bonds. The delocalization index $\delta$ between two bonded atoms indicates the bond order between them.

As expected, QTAIM analysis of Am(CrO$_4$)$_7$Cs$_{11}$ finds eight bond paths terminating at the Am center, one from each of the nearest neighbour O atoms. BCP data for these are given in Table 2, together with the eight $\delta$ values. All of these metrics indicate very ionic Am–O bonding. The $\rho$ values are all well below the 0.1 e/bohr$^3$ value generally taken as the upper limit for an ionic bond, and the significantly positive Laplacian data support this picture. The energy densities are all very close to zero, indicating no covalency, and the $\delta$(Am,O) data average less than 0.3.

In the Introduction, we noted that the enthalpy of complexation of Am$^{III}$ by softer donors can be larger than for Eu$^{III}$. This is sometimes attributed to marginally larger covalency in the 5$f$ complexes. To compare directly the Am–O and Eu–O bonding in our system, we have replaced the Am center with Eu and recomputed the electronic structure and QTAIM metrics; the data are given in Table 2. Consistent with reduced covalency, all of the Eu QTAIM metrics are a little smaller in an absolute sense than their Am counterparts. We stress, however, the highly ionic nature of both Am–O and Eu–O bonding that Table 2 presents; rather than say the lanthanide system is less covalent than the actinide, a better description is that the Eu–O bonds are marginally more ionic than the already highly ionic Am–O analogues.

In summary, the extent or otherwise of Am–O covalency in our Am(CrO$_4$)$_7$Cs$_{11}$ cluster depends on one's definition of the term. The canonical orbitals show extensive mixing between Am-5$f$ and O-2$p$ orbitals, but there is no significant overlap between them. There is thus very little buildup of electron density in the internuclear region, and QTAIM analysis points to a very ionic picture. This view is reinforced by the NBO data, which find six fully localized 5$f$ electrons and no Am–O bonding orbitals. We conclude that there is very little overlap-driven Am–O covalency, though degeneracy-driven covalency is clearly present in the electronic structure. We will now show that this degeneracy-driven covalency plays a major role in the properties of α-CsAm(CrO$_4$)$_2$. Furthermore, the SOC was not included in the cluster calculations presented above; as we are going to see in the next section, our band-structure calculations demonstrate that the SOC substantially influences the electronic structure of this material.

**Band Structure of α-CsM(CrO₄)₂ (M= Sm, Eu, Am).** Here we investigate theoretically the electronic structure of α-CsM(CrO$_4$)$_2$ (M= Sm, Eu, Am) in their respective lattice configurations utilizing LDA+GA and LDA+DMFT, taking fully into account the SOC. This particular selection of *f*-block chromates enables us to investigate how the properties of the M-O chemical bond are affected (i) as M varies between the 4*f* (M=Sm,Eu) and 5*f* series (M=Am), and (ii) as the nominal number of M-4*f* valence electrons varies from 5 (M=Sm) to 6 (M=Am,Eu).

The LDA+DMFT angle-resolved photoemission spectra (ARPES) and the corresponding *f*-electron spectral contributions to the density of states (DOS) are reported in Figure 9. While bare LDA erroneously predicts that α-CsSm(CrO$_4$)$_2$ is a metal and that α-CsAm(CrO$_4$)$_2$ and α-CsEu(CrO$_4$)$_2$ have small band gaps $< 0.7\ eV$, see Figure S3, LDA+DMFT indicates that the materials examined are all insulators with large optical gaps in good agreement with our UV-Vis-NIR absorption experiments. The behavior of the components of *f* spectral weights with single-particle total angular momentum $j = 5/2$ and $j = 7/2$ are significantly different from each other, which indicates that the M-*f* SOC is very strong in all systems. In fact, the atomic spin-orbit splittings of Am, Eu and Sm are about $1.25\ eV$, $0.67\ eV$ and $0.59\ eV$, respectively. Finally, we observe that the ARPES spectra display pronounced incoherent features in all of these materials, as the *f*-electron spectral weight is spread over a broad range of energies.

Both the large enhancement of the band gaps with respect to LDA and the incoherent features of the *f*-electron spectra constitute unequivocal evidence of the strong M-*f* electron correlations in all of these materials. However, the LDA+DMFT electronic structures of these systems are very different from each other. The most significant differences are the following: (I) The $j = 5/2$ component of the self energy $\Sigma_{5/2}(\omega)$ of α-CsSm(CrO$_4$)$_2$ displays a sharp pole, see the right panels of Figure 9, which opens a Mott gap in the LDA spectra. Instead, in α-CsAm(CrO$_4$)$_2$ and α-CsEu(CrO$_4$)$_2$ a gap is already present in the LDA band structure (because of the SOC). (II) The M-*f* degrees of freedom and the O-2*p* bands (below the Fermi level) are hybridized in all of the materials considered, which indicates that the nature of the M–O bonding is never purely ionic. However, the hybridization effects between the M-*f* and the O-2*p* bands are considerably more pronounced in α-CsAm(CrO$_4$)$_2$.

In order to characterize in further detail the nature of the M–O chemical bond and the role of the strong *f*-electron correlations, we consider the local reduced density matrix of the M-*f* electrons $\hat{\rho}_f$, which is formally obtained from the ground state wavefunction of the solid by tracing out all degrees of freedom except the *f* valence shell of one of the Am atoms in the crystal. For this purpose, we conveniently utilize the LDA+GA approach. Let us represent $\hat{\rho}_f$ as:

$$\hat{\rho}_f = \sum_i w_i \hat{r}_i, \qquad [2]$$

where $\hat{r}_i = \hat{P}_i/\text{Tr}[\hat{P}_i]$, $\hat{P}_i$ are projectors over the eigenspaces $V_i$ of $\hat{\rho}_f$, and the probability weights $w_i$ are sorted in descending order. The eigenspaces $V_i$ have well-defined electron occupation $N_i$. Furthermore, since the crystal field splittings are small in this material, the eigenspaces $V_i$ have approximately also a well-defined total angular momentum $J_i$. On the other hand, the orbital angular momentum $L$ and the spin angular momentum $S$ are *not* good quantum numbers, as the SOC of the M-*f* electrons is very strong in all systems. The average number of valence *f*-electrons per M atom is given by $n_f = \text{Tr}[\hat{\rho}_f \hat{N}_f]$, where $\hat{N}_f$ is the corresponding number operator.

The LDA+GA largest local *f*-electron configuration probabilities $w_i$ of our systems and the corresponding quantum numbers $N_i$ and $J_i$ are reported in Tables 3 to 5. According to our calculations, $n_f \sim 6.02$ in α-CsAm(CrO$_4$)$_2$, while $n_f \sim 6.06$ in α-CsEu(CrO$_4$)$_2$, and $n_f \sim 5.07$ in α-CsSm(CrO$_4$)$_2$. As shown in Table 3, the Am-5*f* electronic structure is dominated by a singlet with $N = 6$ electrons and total angular momentum $J = 0$, whose probability weight is $w_0 \sim 0.88$. We note that $\text{Tr}[\hat{r}_0 \hat{S}^2] = \text{Tr}[\hat{r}_0 \hat{L}^2] \sim 2.25 \times (2.25 + 1)$. Similarly, as shown in Table 4, also the Eu-4f electronic structure is dominated by a singlet with $N = 6$ and $J = 0$, where the corresponding probability weight is $w_0 = 0.92$, and $\text{Tr}[\hat{r}_0 \hat{S}^2] = \text{Tr}[\hat{r}_0 \hat{L}^2] \sim 2.8 \times (2.8 + 1)$. Finally, as shown in Table 5, the Sm-4f electronic structure is dominated by a 6-fold degenerate multiplet with $N = 5$ and $J = 5/2$, whose probability weight is $w_0 = 0.93$, $\text{Tr}[\hat{r}_0 \hat{S}^2] \sim 2.42 \times (2.42 + 1)$ and $\text{Tr}[\hat{r}_0 \hat{L}^2] \sim 4.91 \times (4.91 + 1)$.

The reason why the SOC favors a $J = 0$ atomic state in α-CsEu(CrO$_4$)$_2$ and α-CsSm(CrO$_4$)$_2$ is that the nominal number of $f$ valence electrons in Am$^{III}$ and Eu$^{III}$ is 6, which equals the dimensionality of the $j = 5/2$ manifold. The strong SOC creates significant contamination of the spin and orbital angular momentum quantum numbers in both systems (as pointed out above), and substantially lifts the corresponding degeneracy in favor of a $J = 0$ singlet, in agreement with the Hund's third rule. As a consequence, the strong electron correlations do not lead to the formation of a local moment in these materials, which are, in fact, strongly-correlated band insulators. The electronic structure of α-CsSm(CrO$_4$)$_2$ is qualitatively different from the other systems examined. In fact, since the nominal number of valence electrons in Sm$^{III}$ is 5, the Sm-4f $j = 5/2$ manifold is only partially occupied, which allows the strong electron correlation to produce the Mott state.

Interestingly, the total probability weight arising from other multiplets besides $w_0$ is non-negligible for all of the materials examined. The underlying charge fluctuations between the $f$ degrees of freedom and their environment constitute a clear signature of the fact that the $f$ electrons contribution to the bonding is never purely ionic. However, charge fluctuations are particularly pronounced in α-CsAm(CrO$_4$)$_2$, where $1 - w_0 > 10\%$.

In order to more precisely quantify the importance of the covalent contribution to the M–O bonds in the materials considered, it is insightful to compare the physical ground-state energy of the system with the energy minimum realizable in a generic trial quantum state such that the M-*f* valence shell hosts exactly the nominal value of electrons (6 electrons for M=Am, Eu, 5 electrons for M=Sm) entirely disentangled from the rest of the system, i.e., a trial state such that the covalent contribution to the M–O bond is exactly 0, by construction. Formally, such a trial state, which can be easily constructed within the LDA+GA framework, is realized by an electron many-body wavefunction of the form $|\Psi_{ion}\rangle = |\Psi_f\rangle \otimes |\Psi_{env}\rangle$, where $|\Psi_f\rangle$ is the tensor product of all isolated M-*f* atomic states in the lattice, while $|\Psi_{env}\rangle$ is the most general wavefunction of the rest of the system. Because of the variational principle, minimizing the total energy with respect to the most general $|\Psi_{ion}\rangle$ provides us with an energy higher with respect to the physical ground state by a value $\Delta E_{cov} > 0$ per unit cell, which constitutes an unbiased measure of *f*-electron covalency. According to our calculations, $\Delta E_{cov} \sim 0.76\ eV$ for α-CsEu(CrO$_4$)$_2$ and $\Delta E_{cov} \sim 0.67\ eV$ for α-CsSm(CrO$_4$)$_2$, while $\Delta E_{cov} \sim 1.85\ eV$ for α-CsAm(CrO$_4$)$_2$. These large energies indicate that the *f*-electron covalency contributions to the M–O chemical bonds are non-negligible in all of the materials examined. However, from the numerical values of $\Delta E_{cov}$ it clearly emerges that the covalency effects are significantly more important in α-CsAm(CrO$_4$)$_2$ with respect to the other systems.

**Conclusions**

We have prepared and characterized a series of *f*-block chromates, CsM(CrO$_4$)$_2$ (M = La, Pr, Nd, Sm, Eu; Am), and noted pronounced differences between the Am[III] derivative and its lanthanide analogs. In order to investigate the origin of these differences, we have theoretically analyzed the electronic structure of α-CsM(CrO$_4$)$_2$ (M= Sm, Eu, Am) utilizing cluster hybrid DFT and periodic LDA+GA and LDA+DMFT simulations. This particular selection of *f*-block chromates enabled us to investigate how the properties of the M-O chemical bonds are affected (i) as M varies between the 4*f* (M=Sm,Eu) and 5$f$ series (M=Am), and (ii) as the nominal number of M-4$f$ valence electrons varies from 5 (M=Sm) to 6 (M=Am,Eu). Our analysis demonstrates that the $f$-electron correlations are very strong in all of these $f$-block chromates, but the electronic structures and the M–O chemical bonds of these materials are very different from each other: (I) α-CsSm(CrO$_4$)$_2$ is a selective Mott insulator in the $j = 5/2$ manifold. (II) In α-CsAm(CrO$_4$)$_2$ and α-CsEu(CrO$_4$)$_2$ the electron correlations do *not* lead to the formation of a local moment, as the SOC favors a singlet atomic ground state with total angular momentum $J = 0$, in agreement with Hund's third rule. Thus, neither α-CsAm(CrO$_4$)$_2$ nor α-CsEu(CrO$_4$)$_2$ qualify as Mott insulators, but as strongly correlated band insulators, where the occupied $f$-electron spectral weight has mostly $j = 5/2$ character. All systems display hybridization between the M-$f$ and O-2$p$ degrees of freedom, as the nature of the M–O chemical bonds always have a non-negligible covalent component. However, our results indicate that the $f$-electron covalency effects are significantly more pronounced in α-CsAm(CrO$_4$)$_2$ with respect to the other systems examined. Interestingly, we also observed that the covalency effects in the Am$^{III}$ compounds are not present because of significant orbital overlap, but rather because of the degeneracy of the Am$^{III}$ and oxygen 5$f$ and 2$p$ orbitals, and because the large SOC prevents the formation of an Am-5$f$ local moment.

## ▪ ASSOCIATED CONTENT

Supporting Information: Selected crystallographic data, Cartesian coordinates used in the hybrid DFT calculations, magnetic susceptibility data, and CIF files. This material is available free of charge via the Internet at http://pubs.acs.org.

■ **AUTHOR INFORMATION**

**Corresponding Authors** *E-mail: albrecht-schmitt@chem.fsu.edu, nikolas.kaltsoyannis@manchester.ac.uk, lanata@magnet.fsu.edu . Notes: The authors declare no competing financial interest.

**Acknowledgments**

This work was supported as part of the Center for Actinide Science and Technology (CAST), an Energy Frontier Research Center funded by the U.S. Department of Energy, Office of Science, Basic Energy Sciences under Award Number DE-SC0016568. The $^{243}$Am was provided to Florida State University by the Isotope Development and Production for Research and Applications Program through the Radiochemical Engineering and Development Center at Oak Ridge National Laboratory. N.L., T.H. and V.D. were partially supported by the NSF grant DMR-1410132 and the National High Magnetic Field Laboratory. Y.Y. was supported by the U.S. Department of Energy, Office of Science, Basic Energy Sciences, as part of the Computational Materials Science Program. X.D. was supported by the NSF grant DMR-1308141. We are grateful to the University of Manchester's Computational Shared Facility for computational resources and associated support. Part of the calculations were performed utilizing the Extreme Science and Engineering Discovery Environment (XSEDE) by NSF under Grants No. DMR170121.

**References**

(1) Johansson, B.; Rosengren, A. *Phys. Rev. B*. **1975,** 11, 2836.

(2) Smith, J. L.; Haire, R. G. *Science*. **1978,** 200, 535-537.

(3) Janoschek, M.; Das, P.; Chakrabarti, B.; Abernathy, D. L.; Lumsden, M. D.; Lawrence, J. M.; Thompson, J. D.; Lander, G. H.; Mitchell, J. N.; Richmond, S. *Sci. Adv.* **2015,** 1, e1500188.

(4) Janoschek, M.; Lander, G.; Lawrence, J. M.; Bauer, E.; Lashley, J. C.; Lumsden, M.; Abernathy, D. L.; Thompson, J. *Proc. Natl. Acad. Sci. U.S.A.* **2017**, *114*, E268.

(5) Söderlind, P.; Wills, J.; Johansson, B.; Eriksson, O. *Phys. Rev. B*. **1997**, *55*, 1997-2004.

(6) Karraker, D. G.; Stone, J. A.; Jones Jr, E. R.; Edelstein, N. *J. Am. Chem. Soc.* **1970**, *92*, 4841-4845.

(7) Minasian, S. G.; Keith, J. M.; Batista, E. R.; Boland, K. S.; Clark, D. L.; Kozimor, S. A.; Martin, R. L.; Shuh, D. K.; Tyliszczak, T. *Chem. Sci.* **2014**, *5*, 351-359.

(8) Gregson, M.; Lu, E.; Tuna, F.; McInnes, E. J.; Hennig, C.; Scheinost, A. C.; McMaster, J.; Lewis, W.; Blake, A. J.; Kerridge, A. *Chem. Sci.* **2016**, *7*, 3286-3297.

(9) Silver, M. A.; Cary, S. K.; Stritzinger, J. T.; Parker, T. G.; Maron, L.; Albrecht-Schmitt, T. E. *Inorg. Chem.* **2016**, *55*, 5092-5094.

(10) Adam, C.; Kaden, P.; Beele, B. B.; Müllich, U.; Trumm, S.; Geist, A.; Panak, P. J.; Denecke, M. A. *Dalton Trans.* **2013**, *42*, 14068-14074.

(11) Schnaars, D. D.; Gaunt, A. J.; Hayton, T. W.; Jones, M. B.; Kirker, I.; Kaltsoyannis, N.; May, I.; Reilly, S. D.; Scott, B. L.; Wu, G. *Inorg. Chem.* **2012**, *51*, 8557-8566.

(12) Dutkiewicz, M. S.; Farnaby, J. H.; Apostolidis, C.; Colineau, E.; Walter, O.; Magnani, N.; Gardiner, M. G.; Love, J. B.; Kaltsoyannis, N.; Caciuffo, R. *Nat. Chem.* **2016**, *8*, 797-802.

(13) Minasian, S. G.; Keith, J. M.; Batista, E. R.; Boland, K. S.; Clark, D. L.; Conradson, S. D.; Kozimor, S. A.; Martin, R. L.; Schwarz, D. E.; Shuh, D. K. *J. Am. Chem. Soc.* **2012**, *134*, 5586-5597.

(14) Nugent, L.; Baybarz, R.; Burnett, J.; Ryan, J. *J. Phys. Chem.* **1973**, *77*, 1528-1539.

(15) Crosswhite, H.; Crosswhite, H.; Carnall, W.; Paszek, A. *J. Chem. Phys.* **1980**, *72*, 5103-5117.

(16) Sato, T.; Asai, M.; Borschevsky, A.; Stora, T.; Sato, N.; Kaneya, Y.; Tsukada, K.; Düllmann, C. E.; Eberhardt, K.; Eliav, E. *Nature* **2015**, *520*, 209-211.

(17) Lukens, W. W.; Edelstein, N. M.; Magnani, N.; Hayton, T. W.; Fortier, S.; Seaman, L. A. *J. Am. Chem. Soc.* **2013**, *135*, 10742-10754.

(18) Verma, P.; Autschbach, J. *J. Chem. Theory Comput.* **2013**, *9*, 1052-1067.

(19) Silver, M. A.; Cary, S. K.; Johnson, J. A.; Baumbach, R. E.; Arico, A. A.; Luckey, M.; Urban, M.; Wang, J. C.; Polinski, M. J.; Chemey, A. *Science*. **2016**, *353*, 888-894.

(20) Cross, J. N.; Villa, E. M.; Wang, S.; Diwu, J.; Polinski, M. J.; Albrecht-Schmitt, T. E. *Inorg. Chem.* **2012**, *51*, 8419-8424.

(21) Xu, J.; Radkov, E.; Ziegler, M.; Raymond, K. N. *Inorg. Chem.* **2000**, *39*, 4156-4164.

(22) Wilson, R. E. *Inorg. Chem.* **2011**, *50*, 5663-5670.

(23) Diwu, J.; Nelson, A.-G. D.; Albrecht-Schmitt, T. E. *Comments Inorg. Chem.* **2010**, *31*, 46-62.

(24) Diwu, J.; Grant, D. J.; Wang, S.; Gagliardi, L.; Albrecht-Schmitt, T. E. *Inorg. Chem.* **2012**, *51*, 6906-6915.

(25) Diwu, J.; Nelson, A.-G. D.; Wang, S.; Campana, C. F.; Albrecht-Schmitt, T. E. *Inorg. Chem.* **2010**, *49*, 3337-3342.

(26) Cross, J. N.; Duncan, P. M.; Villa, E. M.; Polinski, M. J.; Babo, J.-M.; Alekseev, E. V.; Booth, C. H.; Albrecht-Schmitt, T. *J. Am. Chem. Soc*. **2013**, *135*, 2769-2775.

(27) Polinski, M. J.; Wang, S.; Alekseev, E. V.; Depmeier, W.; Albrecht-Schmitt, T. E. *Angew. Chem. Int. Ed.* **2011**, *50*, 8891-8894.

(28) Silver, M. A.; Albrecht-Schmitt, T. E. *Coord. Chem. Rev.* **2016**, *323*, 36-51.

(29) Schott, J.; Kretzschmar, J.; Acker, M.; Eidner, S.; Kumke, M. U.; Drobot, B.; Barkleit, A.; Taut, S.; Brendler, V.; Stumpf, T. *Dalton Trans.* **2014**, *43*, 11516-11528.

(30) Cary, S. K.; Galley, S. S.; Marsh, M. L.; Hobart, D. L.; Baumbach, R. E.; Cross, J. N.; Stritzinger, J. T.; Polinski, M. J.; Maron, L.; Albrecht-Schmitt, T. E. *Nat. Chem.* **2017**, *9*, 856-861.

(31) Cary, S. K.; Ferrier, M. G.; Baumbach, R. E.; Silver, M. A.; Lezama Pacheco, J.; Kozimor, S. A.; La Pierre, H. S.; Stein, B. W.; Arico, A. A.; Gray, D. L. *Inorg. Chem.* **2016**, *55*, 4373-4380.

(32) Xu, J.; Durbin, P. W.; Kullgren, B.; Ebbe, S. N.; Uhlir, L. C.; Raymond, K. N. *J. Med. Chem.* **2002**, *45*, 3963-3971.

(33) Sturzbecher-Hoehne, M.; Choi, T. A.; Abergel, R. J. *Inorg. Chem.* **2015**, *54*, 3462-3468.

(34) Szigethy, G.; Xu, J.; Gorden, A. E.; Teat, S. J.; Shuh, D. K.; Raymond, K. N. *Eur. J. Inorg. Chem.* **2008**, *13*, 2143-2147.

(35) Miguirditchian, M.; Guillaneux, D.; Guillaumont, D.; Moisy, P.; Madic, C.; Jensen, M. P.; Nash, K. L. *Inorg. Chem.* **2005**, *44*, 1404-1412.

(36) Jensen, M. P.; Bond, A. H. *J. Am. Chem. Soc.* **2002**, *124*, 9870-9877.

(37) Wang, J.; Su, D.; Wang, D.; Ding, S.; Huang, C.; Huang, H.; Hu, X.; Wang, Z.; Li, S. *Inorg. Chem.* **2015**, *54*, 10648-10655.

(38) Martel, L.; Magnani, N.; Vigier, J.-F.; Boshoven, J.; Selfslag, C.; Farnan, I.; Griveau, J.-C.; Somers, J.; Fanghänel, T. *Inorg. Chem.* **2014**, *53*, 6928-6933.

(39) Modolo, G.; Kluxen, P.; Geist, A. *Radiochim. Acta.* **2010**, *98*, 193-201.

(40) Modolo, G.; Wilden, A.; Geist, A.; Magnusson, D.; Malmbeck, R. *Radiochim. Acta.* **2012**, *100*, 715-725.

(41) Gaunt, A. J.; Reilly, S. D.; Enriquez, A. E.; Scott, B. L.; Ibers, J. A.; Sekar, P.; Ingram, K. I.; Kaltsoyannis, N.; Neu, M. P. *Inorg. Chem.* **2008**, *47*, 29-41.

(42) Marsh, M. L.; Albrecht-Schmitt, T. *Dalton Trans.* **2017**, *46*, 9316-9333.

(43) Cary, S. K.; Silver, M. A.; Liu, G.; Wang, J. C.; Bogart, J. A.; Stritzinger, J. T.; Arico, A. A.; Hanson, K.; Schelter, E. J.; Albrecht-Schmitt, T. E. *Inorg. Chem.* **2015**, *54*, 11399-11404.

(44) Cary, S. K.; Vasiliu, M.; Baumbach, R. E.; Stritzinger, J. T.; Green, T. D.; Diefenbach, K.; Cross, J. N.; Knappenberger, K. L.; Liu, G.; Silver, M. A. *Nat. Commun.* **2015**, *6*, 6827-6834.

(45) Polinski, M. J.; Iii, E. B. G.; Maurice, R.; Planas, N.; Stritzinger, J. T.; Parker, T. G.; Cross, J. N.; Green, T. D.; Alekseev, E. V.; Van Cleve, S. M. *Nat. Chem.* **2014**, *6*, 387-392.

(46) Dupouy, G.; Bonhoure, I.; Conradson, S. D.; Dumas, T.; Hennig, C.; Le Naour, C.; Moisy, P.; Petit, S.; Scheinost, A. C.; Simoni, E. *Eur. J. Inorg. Chem.* **2011**, *10*, 1560-1569.

(47) Wall, T. F.; Jan, S.; Autillo, M.; Nash, K. L.; Guerin, L.; Naour, C. L.; Moisy, P.; Berthon, C. *Inorg. Chem.* **2014**, 10, 2450-2459.

(48) Leguay, S. b.; Vercouter, T.; Topin, S.; Aupiais, J.; Guillaumont, D.; Miguirditchian, M.; Moisy, P.; Le Naour, C. *Inorg. Chem.* **2012**, *51*, 12638-12649.

(49) Dau, P. D.; Shuh, D. K.; Sturzbecher-Hoehne, M.; Abergel, R. J.; Gibson, J. K. *Dalton Trans.* **2016**, *45*, 12338-12345.

(50) Deblonde, G. J.-P.; Sturzbecher-Hoehne, M.; Rupert, P. B.; An, D. D.; Illy, M.-C.; Ralston, C. Y.; Brabec, J.; de Jong, W. A.; Strong, R. K.; Abergel, R. J. *Nat. Chem.* **2017**, *9*, 843-849.

(51) MacDonald, M. R.; Bates, J. E.; Ziller, J. W.; Furche, F.; Evans, W. J. *J. Am. Chem. Soc.* **2013**, *135*, 9857-9868.

(52) Fieser, M. E.; MacDonald, M. R.; Krull, B. T.; Bates, J. E.; Ziller, J. W.; Furche, F.; Evans, W. J. *J. Am. Chem. Soc.* **2014**, *137*, 369-382.

(53) Windorff, C. J.; MacDonald, M. R.; Meihaus, K. R.; Ziller, J. W.; Long, J. R.; Evans, W. J. *Chem. Eur. J.* **2016**, *22*, 772-782.

(54) Windorff, C. J.; Chen, G. P.; Cross, J. N.; Evans, W. J.; Furche, F.; Gaunt, A. J.; Janicke, M. T.; Kozimor, S. A.; Scott, B. L. *J. Am. Chem. Soc.* **2017**, *139*, 3970-3973.

(55) Kelley, M. P.; Su, J.; Urban, M.; Luckey, M.; Batista, E. R.; Yang, P.; Shafer, J. C. *J. Am. Chem. Soc.* **2017**, *139*, 9901-9908.

(56) Prodan, I. D.; Scuseria, G. E.; Martin, R. L. *Phys. Rev. B* **2007**, *76*, 033101.

(57) Neidig, M. L.; Clark, D. L.; Martin, R. L. *Coord. Chem. Rev.* **2013**, *257*, 394-406.

(58) Choppin, G.; Morgenstern, A. *J. Radioanal. Nucl. Chem.* **2000**, *243*, 45-51.

(59) Lukens, W. W.; Magnani, N.; Tyliszczak, T.; Pearce, C. I.; Shuh, D. K. *Environ. Sci. Technol.* **2016**, *50*, 13160-13168.

(60) Krivovichev, S.; Burns, P.; Tananaev, I. *Structural chemistry of inorganic actinide compounds*; Elsevier, 2006.

(61) Sullens, T. A.; Albrecht-Schmitt, T. E. *Acta Crystallogr. Sect. Sect. E: Struct. Rep. Online*. **2006,** *62*, i258-i260.

(62) Sigmon, G. E.; Burns, P. C. *J. Solid State Chem.* **2010,** *183*, 1604-1608.

(63) Krivovichev, S. V.; Burns, P. C. *Z. Anorg. Allg. Chem.* **2003**, *629*, 1965-1968.

(64) Siidra, O. I.; Nazarchuk, E. V.; Krivovichev, S. V. *Z. Anorg. Allg. Chem.* **2012**, *638*, 976-981.

(65) Siidra, O. I.; Nazarchuk, E. V.; Krivovichev, S. V. *Eur. J. Inorg. Chem.* **2012**, *2*, 194-197.

(66) Siidra, O. I.; Nazarchuk, E. V.; Krivovichev, S. V. *Z. Anorg. Allg. Chem.* **2012**, *638*, 982-986.

(67) Fedoseev, A.; Budantseva, N.; Grigor'ev, M.; Perminov, V. *Radiokhimiya*. **1991**, *33*, 7-19.

(68) Charushnikova, I.; Fedoseev, A. *Radiochemistry*. **2013**, *55*, 11-15.

(69) Melnikov, P.; Parada, C.; Bueno, I.; Moran, E. *J. Alloys Compd.* **1993**, *190*, 265-267.

(70) Melnikov, P.; Bueno, I.; Parada, C.; Morán, E.; León, C.; Santamaría, J.; Sánchez-Quesada, F. *Solid State Ionics*. **1993**, *63*, 581-584.

(71) Melnikov, P.; Ferracin, L. *J. Alloys Compd.* **1995**, *224*, L5-L6.

(72) Sheldrick, G. M. *Acta Cryst. A*. **2015**, *71*, 3-8.

(73) Spek, A. L. *Acta Crystallogr. Sect. D. Biol. Crystallogr.* **2009**, *65*, 148-155.

(74) Frisch, M.; Trucks, G.; Schlegel, H. B.; Scuseria, G.; Robb, M.; Cheeseman, J.; Scalmani, G.; Barone, V.; Mennucci, B.; Petersson, G. *et al.*; Gaussian, Inc., Wallingford CT, 2009.

(75) Cao, X.; Dolg, M. *J. Mol. Struc.-THEOCHEM*. **2004**, *673*, 203-209.

(76) Cao, X.; Dolg, M.; Stoll, H. *J. Chem. Phys.* **2003**, *118*, 487-496.

(77) Leininger, T.; Nicklass, A.; Küchle, W.; Stoll, H.; Dolg, M.; Bergner, A. *Chem. Phys. Lett.* **1996**, *255*, 274-280.

(78) Adamo, C.; Barone, V. *J. Chem. Phys.* **1999**, *110*, 6158-6170.

(79) Keith, T. A. AIMAll v 16.01.09. http://aim.tkgristmill.com, **2015**.

(80) Glendening, E.; Badenhoop, J.; Reed, A.; Carpenter, J.; Bohmann, J.; Morales, C.; Landis, C.; Weinhold, F. In *Theoretical Chemistry Institute, University of Wisconsin, Madison, WI*, 2013.

(81) Gunnarsson, O.; Lundqvist, B. I. *Phys. Rev. B*. **1976**, *13*, 4274.

(82) Georges, A.; Kotliar, G.; Krauth, W.; Rozenberg, M. J. *Rev. Mod. Phys.* **1996**, *68*, 13.

(83) Anisimov, V.; Poteryaev, A.; Korotin, M.; Anokhin, A.; Kotliar, G. *J. Phys.: Condens. Matter*. **1997**, *9*, 7359.

(84) Lichtenstein, A.; Katsnelson, M. *Phys. Rev. B*. **2000**, *62*, R9283.

(85) Gutzwiller, M. C. *Phys. Rev.* **1965**, *137*, A1726.

(86) Deng, X.; Wang, L.; Dai, X.; Fang, Z. *Phys. Rev. B*. **2009**, *79*, 075114.

(87) Lanatà, N.; Yao, Y.; Wang, C.-Z.; Ho, K.-M.; Kotliar, G. *Phys. Rev. X*. **2015**, *5*, 011008.

(88) Lanatà, N.; Yao, Y.; Deng, X.; Dobrosavljević, V.; Kotliar, G. *Phys. Rev. Lett.* **2017**, *118*, 126401.

(89) Blaha, P.; Schwarz, K.; Madsen, G.; Kvasnicka, D.; Luitz, J. (Karlheinz Schwarz, Techn. Universität Wien, Austria, 2001).

(90) Haule, K.; Yee, C.-H.; Kim, K. *Phys. Rev. B*. **2010**, *81*, 195107.

(91) Werner, P.; Comanac, A.; De'Medici, L.; Troyer, M.; Millis, A. J. *Phys. Rev. Lett.* **2006**, *97*, 076405.

(92) Ruiz-Martínez, A.; Casanova, D.; Alvarez, S. *Chem. Eur. J.* **2008**, *14*, 1291-1303.

(93) Kuzina, T.; Shakhno, I.; Krachak, A.; Plyushchev, V. *Zhurnal Neorganicheskoj Khimii*. **1973**, *18*, 2727-2730.

(94) Bueno, I.; Parada, C.; García, O.; Puebla, E. G.; Monge, A.; Valero, C. R. *J. Chem. Soc., Dalton Trans.* **1988**, 1911-1914.

(95) Bueno, I.; Garcia, O.; Parada, C.; Puche, R. S. *J. Less-Common MET*. **1988**, *139*, 261-271.

(96) Bueno, I.; Parada, C.; Puche, R. S.; Botto, I.; Baran, E. *J. Phys. Chem. Solids*. **1990**, *51*, 1117-1121.

(97) Bueno, I.; Parada, C.; Hermoso, J.; Vegas, A.; Marti, M. *J. Solid State Chem.* **1990**, *85*, 83-87.

(98) Runde, W.; Bean, A. C.; Brodnax, L. F.; Scott, B. L. *Inorg. Chem.* **2006**, *45*, 2479-2482.

(99) Sykora, R. E.; Assefa, Z.; Haire, R. G.; Albrecht-Schmitt, T. E. *Inorg. Chem.* **2006**, *45*, 475-477.

(100) Sykora, R. E.; Assefa, Z.; Haire, R. G.; Albrecht-Schmitt, T. E. *J. Solid State Chem.* **2004**, *177*, 4413-4419.

(101) Douglas, P.; Hector, A. L.; Levason, W.; Light, M. E.; Matthews, M. L.; Webster, M. *Z. Anorg. Allg. Chem.* **2004**, *630*, 479-483.

(102) Nassau, K.; Shiever, J.; Prescott, B. *J. Solid State Chem.* **1975**, *14*, 122-132.

(103) Nassau, K.; Shiever, J.; Prescott, B.; Cooper, A. *J. Solid State Chem.* **1974**, *11*, 314-318.

(104) Abrahams, S.; Bernstein, J.; Nassau, K. *J. Solid State Chem.* **1976**, *16*, 173-184.

(105) Abrahams, S.; Sherwood, R.; Bernstein, J.; Nassau, K. *J. Solid State Chem.* **1973**, *7*, 205-212.

(106) Bentria, B.; Benbertal, D.; Bagieu-Beucher, M.; Masse, R.; Mosset, A. *J. Chem. Crystallogr.* **2003**, *33*, 867-873.

(107) Hector, A. L.; Henderson, S. J.; Levason, W.; Webster, M. *Z. Anorg. Allg. Chem.* **2002**, *628*, 198-202.

(108) Phanon, D.; Mosset, A.; Gautier-Luneau, I. *Solid State Sci.* **2007**, *9*, 496-505.

(109) Wybourne, B. G. *Spectroscopic properties of rare earths*; Interscience New York, 1965.

(110) Sykora, R. E.; Assefa, Z.; Haire, R. G.; Albrecht-Schmitt, T. E. *Inorg. Chem.* **2005**, *44*, 5667-5676.

(111) Neidig, M. L.; Clark, D. L.; Martin, R. L. *Coord. Chem. Rev.* **2013**, *257*, 394-406.

(112) Kaltsoyannis, N. *Inorg. Chem.* **2012**, *52*, 3407-3413.

(113) Tassell, M. J.; Kaltsoyannis, N. *Dalton Trans.* **2010**, *39*, 6719-6725.

(114) Kirker, I.; Kaltsoyannis, N. *Dalton Trans.* **2011**, *40*, 124-131.

(115) Jones, M. B.; Gaunt, A. J.; Gordon, J. C.; Kaltsoyannis, N.; Neu, M. P.; Scott, B. L. *Chem. Sci.* **2013**, *4*, 1189-1203.

(116) Kaltsoyannis, N. *Dalton Trans.* **2016**, *45*, 3158-3162.

(117) Lanatà, N.; Strand, H. U.; Yao, Y.; Kotliar, G. *Phys. Rev. Lett.* **2014**, *113*, 036402.

(118) Lanata, N.; Yao, Y.-X.; Wang, C.-Z.; Ho, K.-M.; Schmalian, J.; Haule, K.; Kotliar, G. *Phys. Rev. Lett.* **2013**, *111*, 196801.

Table 1: Selected Bond Distances (Å) for CsM(CrO$_4$)$_2$. (M= Nd, Am)

| CsNd(CrO$_4$)$_2$ | | α-CsAm(CrO$_4$)$_2$ | | β−CsAm(CrO$_4$)$_2$ | |
|---|---|---|---|---|---|
| Nd–O(1) | 2.381(2) | Am(1)–O(1) | 2.441(3) | Am(1)–O(2) | 2.354(4) |
| Nd–O(2) | 2.425(2) | Am(1)–O(1) | 2.466(3) | Am(1)–O(3) | 2.408(4) |
| Nd–O(3) | 2.573(2) | Am(1)–O(2) | 2.449(3) | Am(1)–O(4) | 2.554(4) |
| | | Am(1)–O(4) | 2.430(3) | | |
| | | Am(1)–O(4) | 2.635(3) | | |
| | | Am(1)–O(5) | 2.425(4) | | |
| | | Am(1)–O(6) | 2.332(4) | | |
| | | Am(1)–O(8) | 2.324(4) | | |
| Average Nd–O | 2.460(2) | Average Am–O | 2.438(4) | | 2.439(4) |

**Table 2.** QTAIM M–O Bond Critical Point parameters (au) and delocalization indices for Am(CrO$_4$)$_7$Cs$_{11}$ and Eu(CrO$_4$)$_7$Cs$_{11}$. Data for Am in upright text, that for Eu in italics.

|       | $\rho$        | $\nabla^2\rho$ | $H$            | $\delta$(M,O)  |
|-------|---------------|----------------|----------------|----------------|
| M–O13 | 0.051 *0.044* | 0.216 *0.192*  | 0.003 *0.001*  | 0.255 *0.206*  |
| M–O25 | 0.054 *0.048* | 0.224 *0.198*  | 0.004 *-0.001* | 0.289 *0.240*  |
| M–O10 | 0.054 *0.049* | 0.212 *0.186*  | 0.004 *-0.002* | 0.331 *0.280*  |
| M–O35 | 0.068 *0.059* | 0.290 *0.257*  | 0.008 *-0.004* | 0.348 *0.283*  |
| M–O11 | 0.048 *0.043* | 0.204 *0.179*  | 0.002 *0.000*  | 0.268 *0.212*  |
| M–O24 | 0.053 *0.047* | 0.220 *0.195*  | 0.003 *-0.001* | 0.303 *0.251*  |
| M–O30 | 0.067 *0.060* | 0.282 *0.244*  | 0.008 *-0.005* | 0.370 *0.308*  |
| M–O22 | 0.035 *0.030* | 0.137 *0.119*  | 0.000 *0.002*  | 0.209 *0.178*  |

**Table 3.** Parameters of the Am-5$f$ reduced density matrix in CsAm(CrO$_4$), computed employing LDA+GA assuming $U$=6 and $J$=0.7, see Eq. 2. Largest probability weights $w_i$, corresponding quantum labels $N_i$ (number of electrons) and $J_i$ (total angular momentum).

| $i$ | 1 | 2 | 3 | 4 |
|---|---|---|---|---|
| $w_i$ | 0.88 | 0.06 | 0.04 | 0.002 |
| $N_i$ | 6 | 7 | 5 | 6 |
| $J_i$ | 0 | 3.5 | 2.5 | 6 |

**Table 4.** Parameters of the Eu-5$f$ reduced density matrix in CsEu(CrO$_4$), computed employing LDA+GA assuming $U$=6 and $J$=0.7, see Eq. 2. Largest probability weights $w_i$, corresponding quantum labels $N_i$ (number of electrons) and $J_i$ (total angular momentum).

| $i$ | 1 | 2 | 3 | 4 |
|---|---|---|---|---|
| $w_i$ | 0.921 | 0.069 | 0.007 | 0.004 |
| $N_i$ | 6 | 7 | 5 | 6 |
| $J_i$ | 0 | 3.5 | 2.5 | 6 |

**Table 5.** Parameters of the Sm-5$f$ reduced density matrix in CsSm(CrO$_4$), computed employing LDA+GA assuming $U$=6 and $J$=0.7, see Eq. 2. Largest probability weights $w_i$, corresponding quantum labels $N_i$ (number of electrons) and $J_i$ (total angular momentum).

| $i$ | 1 | 2 | 3 | 4 |
|---|---|---|---|---|
| $w_i$ | 0.93 | 0.017 | 0.015 | 0.012 |
| $N_i$ | 5 | 6 | 6 | 6 |
| $J_i$ | 2.5 | 2 | 6 | 3 |

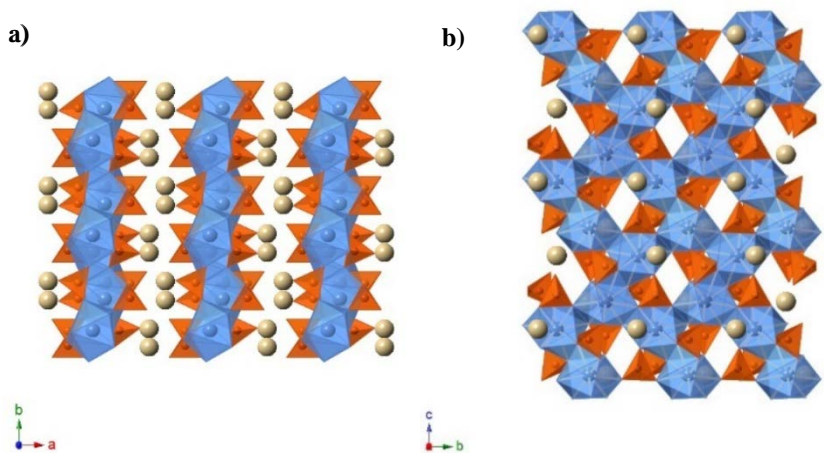

**Figure 1.** Polyhedral representations of CsLn(CrO$_4$)$_2$ (Ln = La, Pr). (a) View along the *c* axis showing stacking of the lanthanum chromate layers with Cs$^+$ cations in the interlayer space. (b) Depiction of part of a single [Ln(CrO$_4$)$_2$]$^{1-}$ layer. Ln$^{III}$ is represented as blue polyhedra, CrO$_4^{2-}$ as orange tetrahedra, and Cs$^+$ as tan spheres.

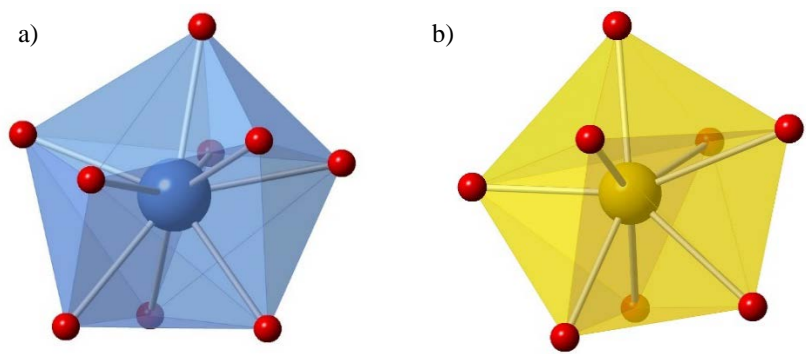

**Figure 2.** a) Nine-coordinate environment of the Ln$^{III}$ cations in CsLn(CrO$_4$)$_2$ (Ln = La, Pr) showing the approximate muffin geometry. b) Eight-coordinate environments Ln$^{III}$ cations in CsLn(CrO$_4$)$_2$ (Ln = Nd, Sm, Eu). This geometry is closest to a trigonal dodecahedron. Oxygen atoms are shown as red spheres.

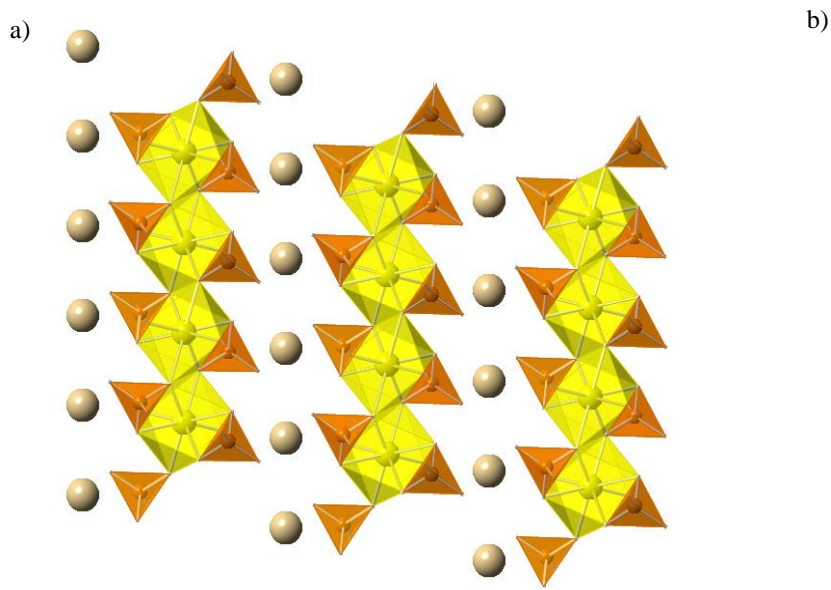

**Figure 3.** Polyhedral representations of the structure of β-CsM(CrO$_4$)$_2$ (Ln = (M = Nd, Sm, Eu; Am) (a) A view along the *b* axis showing the stacking of the layers. (b) Depiction of part of a [M(CrO$_4$)$_2$]$^{1-}$ layer. M$^{III}$ cations are located within the yellow polyhedra, CrO$_4^{2-}$ anions are shown as orange tetrahedra, and Cs$^+$ cations as tan spheres.

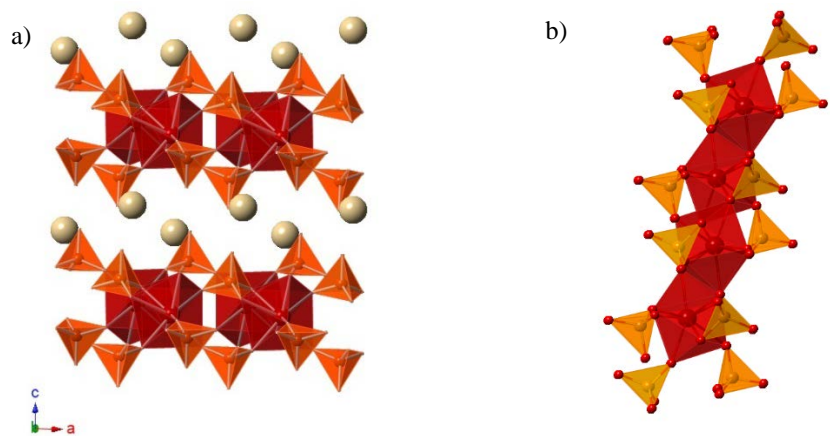

**Figure 4.** Polyhedral representations of α-CsAm(CrO$_4$)$_2$. (a) A view along the *b* axis showing the stacking of the layers. (b) An illustration of a single chain that the layers are formed from. In both views the Am$^{III}$ cations are located within the dark-red polyhedral. CrO$_4^{2-}$ anions are shown as orange tetrahedra, and Cs$^+$ cations as tan spheres.

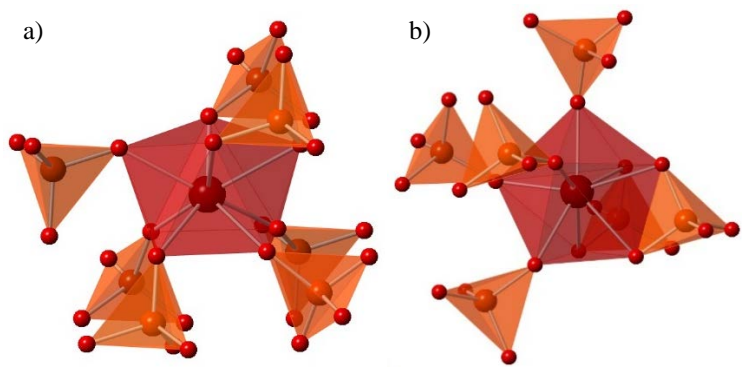

**Figure 5.** Depictions of the local environment of Am$^{III}$ cations in α- (a) and (b) β-CsAm(CrO$_4$)$_2$. In the former case the Am$^{III}$ polyhedron is best approximated by a bicapped trigonal prism ($C_{2v}$); whereas in the latter it is closer to a trigonal dodecahedron ($D_{2d}$). Moreover, in α- there is only one chelating CrO$_4^{2-}$ anion, while in β- there are two.

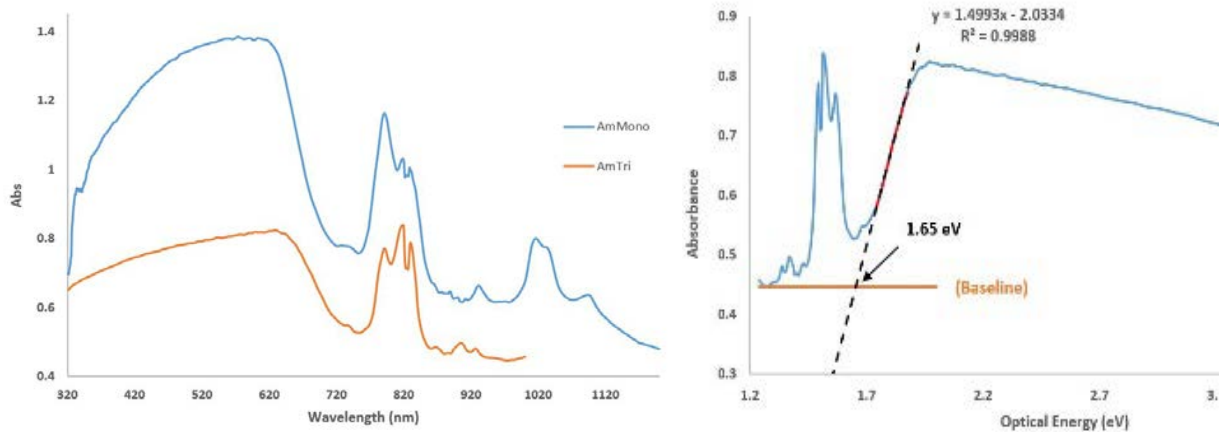

**Figure 6.** a) Solid-state UV-vis-NIR absorption spectrum of α-CsAm(CrO$_4$)$_2$ and β-CsAm(CrO$_4$)$_2$.

b) Absorbance vs. Optical Energy plot of α-CsAm(CrO$_4$)$_2$ showing a band gap of ~1.65 eV.

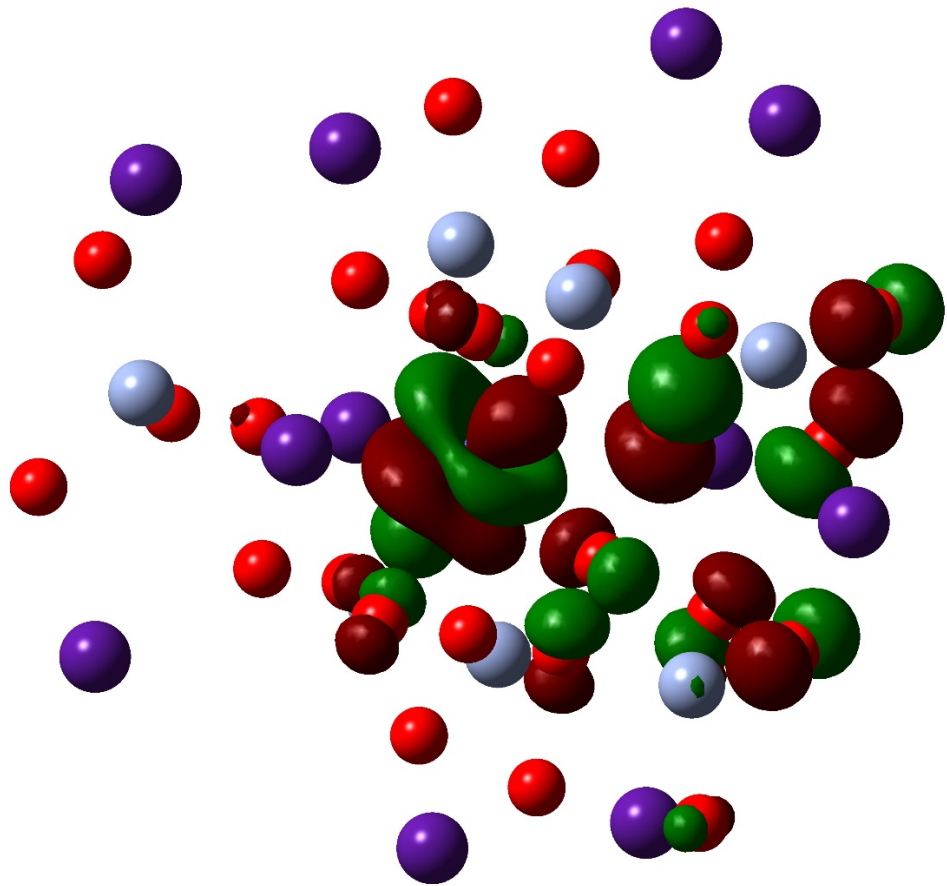

**Figure 7.** MO 266α. Isovalue = 0.035. Dark red and green are the phases of the wavefunction. Grey spheres = Cr, red spheres = O, purple spheres = Cs. The Am atom is in the center of the image. See Table S4 for a detailed breakdown of the atomic orbital contributions.

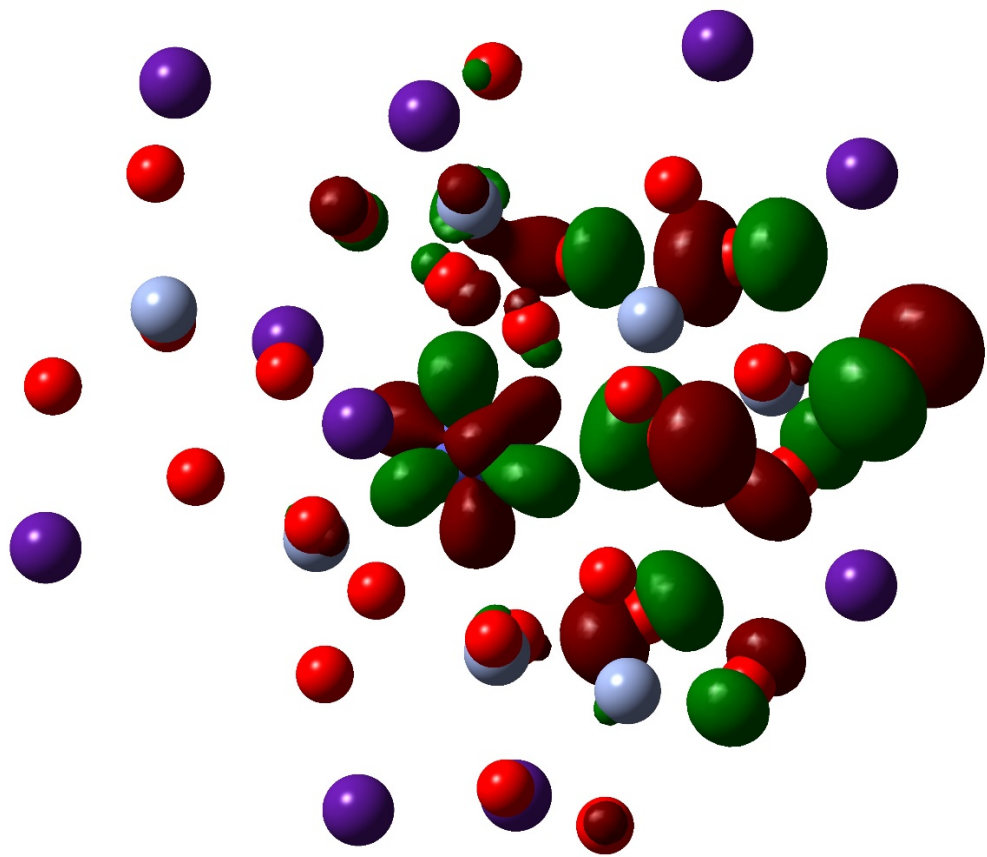

**Figure 8.** MO 262α. Isovalue = 0.035. Dark red and green are the phases of the wavefunction. Grey spheres = Cr, red spheres = O, purple spheres = Cs. The Am atom is in the center of the image. See SI Table 4 for a detailed breakdown of the atomic orbital contributions.

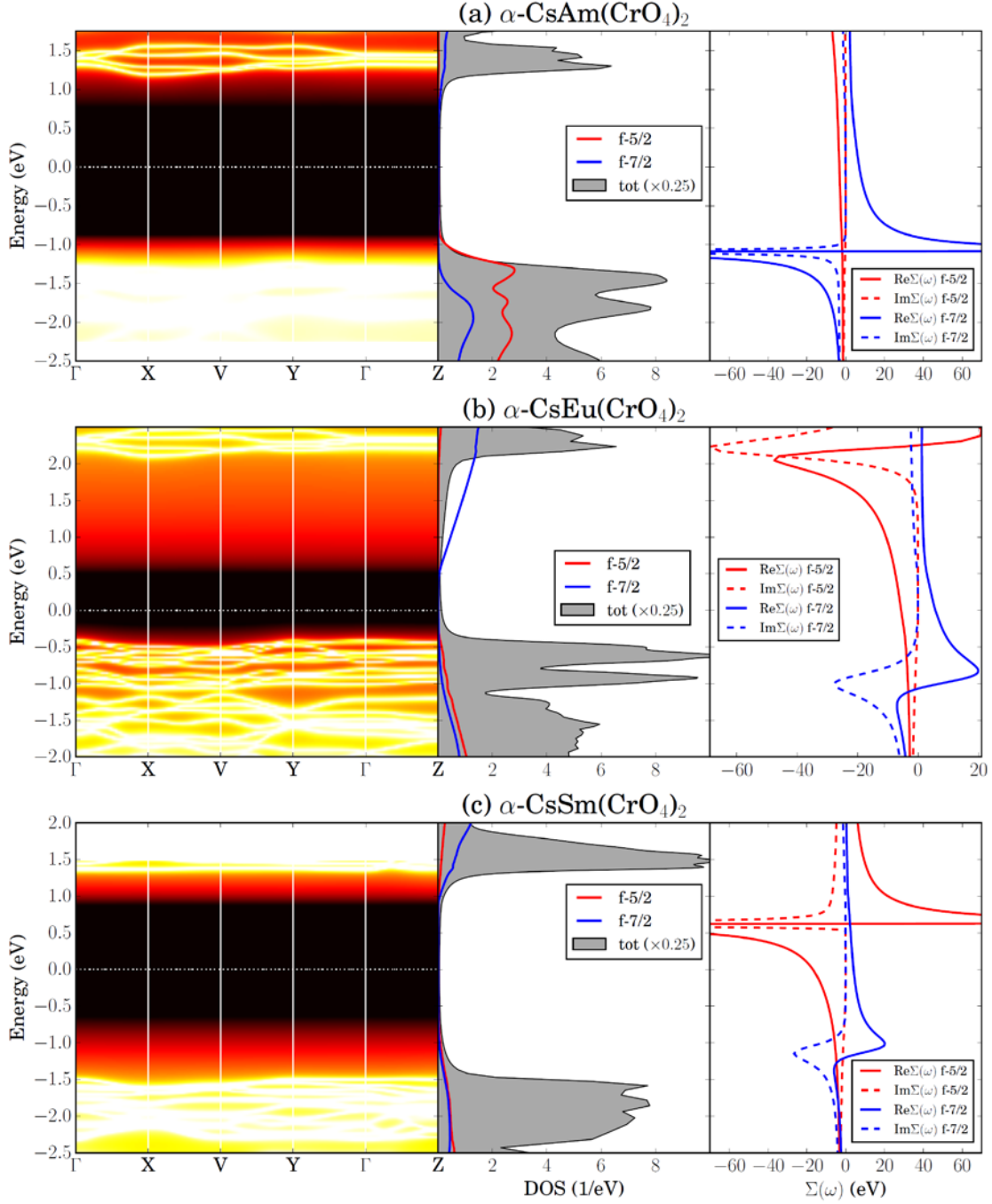

**Figure 9**. LDA+DMFT ARPES spectra computed at $T = 290\ K$ for α-CsAm(CrO$_4$)$_2$, α-CsEu(CrO$_4$)$_2$, and α-CsSm(CrO$_4$)$_2$. The 5/2 and 7/2 *f* spectral contributions to the local DOS and the 5/2 and 7/2 components of the self-energies, $\Sigma(\omega)$ are displayed in the right panels.

**TOC Graphic**

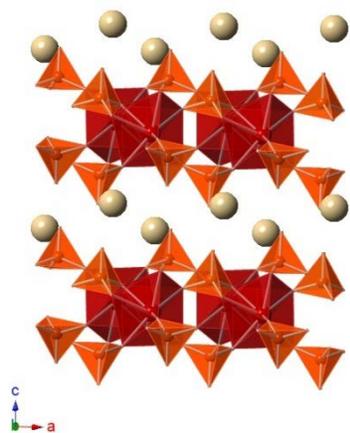 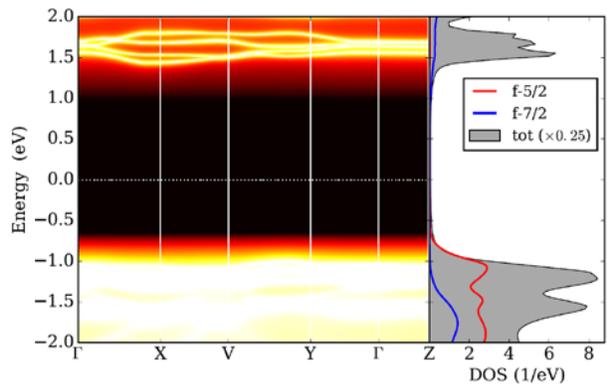